\documentclass[aps,prd,12pt,notitlepage,showpacs,nofootinbib,tightenlines]{revtex4-1}
\usepackage{amsmath}
\usepackage{amssymb}
\usepackage{epsfig}
\usepackage{graphicx}
\usepackage{bm}
\usepackage{times}
\usepackage{braket}
\usepackage{xcolor}
\usepackage{slashed}
\usepackage{hyperref}
\usepackage{multirow}
\usepackage{gensymb}
\usepackage{longtable}
\usepackage{booktabs}
\usepackage{array}
\usepackage{anyfontsize}
\usepackage{ulem}
\usepackage{wrapfig}
\usepackage{cancel}

\begin{document}
\linespread{1}
\setlength{\oddsidemargin}{-5pt}
\setlength{\topmargin}{-0.6in}
\setlength{\textheight}{24cm}
\addtolength{\jot}{5pt}
\addtolength{\arraycolsep}{-3pt}
\renewcommand{\arraystretch}{1.2}
\renewcommand{\textfraction}{0}

\definecolor{blue}{rgb}{0,0,0.8}
\definecolor{red}{rgb}{1,0,0}
\definecolor{purple}{rgb}{0.6,0,0}
\newcommand{\blue}{\color{blue}}
\newcommand{\red}{\color{red}}
\newcommand{\purple}{\color{purple}}
\newcommand{\cheng}[1]{{\blue #1}}
\newcommand{\hnl}[1]{{\red #1}}
\newcommand{\zjx}[1]{{\purple #1}}
\newcommand{\beq}{\begin{eqnarray}}
\newcommand{\eeq}{\end{eqnarray}}
\newcommand{\non}{\nonumber\\ }
\newcommand{\jpsi}{J/\Psi}
\newcommand{\ppa}{\phi_\pi^{\rm A}}
\newcommand{\ppp}{\phi_\pi^{\rm P}}
\newcommand{\ppt}{\phi_\pi^{\rm T}}
\newcommand{\ov}{ \overline }
\newcommand{\zerot}{ {\textbf 0_{\rm T}} }
\newcommand{\kt}{k_{\rm T} }
\newcommand{\fb}{f_{\rm B} }
\newcommand{\fk}{f_{\rm K} }
\newcommand{\rk}{r_{\rm K} }
\newcommand{\mb}{m_{\rm B} }
\newcommand{\mw}{m_{\rm W} }
\newcommand{\im}{{\rm Im} }
\newcommand{\kks}{K^{(*)}}
\newcommand{\acp}{{\cal A}_{\rm CP}}
\newcommand{\pb}{\phi_{\rm B}}
\newcommand{\xeba}{\bar{x}_2}
\newcommand{\xsba}{\bar{x}_3}
\newcommand{\peas}{\phi^A}
\newcommand{\pvsl}{ p \hspace{-2.0truemm}/_{K^*} }
\newcommand{\esl}{ \epsilon \hspace{-2.1truemm}/ }
\newcommand{\psl}{ p \hspace{-2truemm}/ }
\newcommand{\ksl}{ k \hspace{-2.2truemm}/ }
\newcommand{\lsl}{ l \hspace{-2.2truemm}/ }
\newcommand{\nsl}{ n \hspace{-2.2truemm}/ }
\newcommand{\vsl}{ v \hspace{-2.2truemm}/ }
\newcommand{\zsl}{ z \hspace{-2.2truemm}/ }
\newcommand{\epsl}{\epsilon \hspace{-1.8truemm}/\,  }
\newcommand{\bfkk}{{\bf k} }
\newcommand{\calm}{ {\cal M} }
\newcommand{\calh}{ {\cal H} }
\newcommand{\calo}{ {\cal O} }
\newcommand{\calp}{ {\cal P} }
\newcommand{\cald}{ {\cal D} }
\def \appb{{\bf Acta. Phys. Polon. B }  }
\def \cpc{ {\bf Chin. Phys. C } }
\def \ctp{ {\bf Commun. Theor. Phys. } }
\def \epjc{{\bf Eur. Phys. J. C} }
\def \jhep{{\bf J. High Energy Phys. } }
\def \jpg{ {\bf J. Phys. G} }
\def \mpla{{\bf Mod. Phys. Lett. A } }
\def \npb{ {\bf Nucl. Phys. B} }
\def \plb{ {\bf Phys. Lett. B} }
\def \pr{  {\bf Phys. Rep.} }
\def \prc{ {\bf Phys. Rev. C }}
\def \prd{ {\bf Phys. Rev. D} }
\def \prl{ {\bf Phys. Rev. Lett.}  }
\def \ptp{ {\bf Prog. Theor. Phys. }}
\def \zpc{ {\bf Z. Phys. C}  }
\def \jpg{ {\bf J.Phys.-G-}  }
\def \ap{ {\bf Ann. of Phys}  }

\title{Revisiting the factorization theorem for $\rho\gamma^{*} \to \pi(\rho)$ at twist 3}
\author{Shan Cheng$^{\,a}$} \email{cheng@physik.uni-siegen.de}
\author{Ya-lan Zhang$^{\,b}$}
\author{Jun Hua$^{\,b}$}
\author{Hsiang-nan Li$^{\,c}$} \email{hnli@phys.sinica.edu.tw}
\author{Zhen-jun Xiao$^{\,b}$} \email{xiaozhenjun@njnu.edu.cn}

\affiliation{\it
$^a$ Theoretische Elementarteilchenphysik, Naturwissenschaftlich
Technische Fakult$\ddot{a}$t, Universi$\ddot{a}$t Siegen, 57068 Siegen, Germany,\\
$^b$ Department of Physics and Institute of Theoretical Physics,
Nanjing Normal University, Nanjing, Jiangsu 210023, People's Republic of China,\\
$^c$ Institute of Physics, Academia Sinica, Taipei, Taiwan 115, Republic of China}

\date{\today}
\begin{abstract}
We revisit the proof of the perturbative QCD factorization for the
exclusive processes $\rho\gamma^{\star} \to \pi(\rho)$ at the two-parton
twist-3 level. It is pointed out that the residual collinear divergences
observed in the literature, which break the factorization of the above
processes at the considered accuracy, are attributed to the improper
insertion of the Fierz identity for factorizing the fermion flow. We show
that the factorization theorem indeed holds at the two-parton twist-3 level
after the mishandling is corrected.

\end{abstract}
\maketitle

\section{Introduction}
The factorization theorem is the foundation of the perturbative QCD
formalism~\cite{LepageZB,LepageZA,BrodskyNY,EfremovQK,MusatovPU,DuncanHI},
which states that nonperturbative dynamics in a hard QCD process can be
factorized into universal nonlocal hadronic matrix elements
defined in an infinite momentum frame. Recently, some of us have attempted to
extend the proof of the collinear factorization for exclusive processes
involving only pseudoscalar mesons at the subleading-power (twist)
accuracy~\cite{LiHH,NagashimaIW}
to those involving also vector mesons: it was examined whether the collinear
divergences in the light-to-light scattering $\rho \gamma^{\star} \to \pi(\rho)$
are factorized into the two-parton twist-3 $\rho$ meson light-cone
distribution amplitudes~\cite{ChengRRA,ZhangVOR}. Their next-to-leading-order (NLO)
analysis indicated that the triple-gluon vertex gives residual
collinear contributions, which cannot be absorbed into the meson distribution
amplitudes. Namely, the universality of the meson distribution amplitudes, and
thus the collinear factorization for $\rho \gamma^{\star} \to \pi(\rho)$, was
violated at the twist-3 level.

In this paper we revisit the factorization theorem for the above
processes, pointing out that the residual collinear divergences at the
two-parton twist-3 level observed in~\cite{ChengRRA,ZhangVOR} are attributed to
the improper insertion of the Fierz identity:
the twist-3 spin projectors were inserted into the scattering amplitudes first
to factorize the fermion flow between the hadronic matrix elements and the hard
kernels; radiative gluons were added to the hard kernels and the resultant
infrared divergences were investigated subsequently.
A hard kernel is not a physical quantity, based on which the matching between
the full QCD and the effective theory for infrared physics cannot be performed
correctly. For instance, missing the power-law behavior of hadronic matrix
elements would lead to wrong power counting for
scattering amplitudes. This is the reason why some collinear divergences
survive the matching, and break the factorization theorem.
We claim that the appropriate procedure to examine the factorization theorem
at a subleading level follows the one proposed in~\cite{NagashimaIW}, which
starts with an analysis of infrared divergences in quark-level scattering amplitudes.
The Fierz identity is inserted to factorize the fermion flow between the hadronic
matrix elements and the hard kernels, after the collinear divergences have been
absorbed into the former.

We first derive the power counting for the products of various gamma matrices
with a valence quark spinor by means of the equation of motion without the
three-parton terms. It is then demonstrated that
the residual collinear divergences in radiative corrections to the
$\rho \gamma^{\star} \to \pi(\rho)$ scattering amplitudes are actually
power suppressed and negligible up to twist 3. As a result,
the soft divergences cancel, and the collinear divergences can be absorbed
completely into the hadronic matrix elements. The twist-3 spin projectors are
inserted into the scattering amplitudes at this stage to factorize the
fermion flow between the hadronic matrix elements and the hard
kernels~\cite{LiHH,NagashimaIW}, with the former defining the two-parton twist-3
meson distribution amplitudes. After proving the collinear factorization, we allow
valence quarks to be off their mass shell by including quark transverse momenta
$k_T$. The collinear divergences, regularized into $\ln k_T$ in this formalism,
can be collected into the transverse-momentum-dependent (TMD) two-parton twist-3
meson wave functions in a similar way. It is concluded, contrary to the observation
in~\cite{ChengRRA,ZhangVOR}, that the factorization theorem indeed holds for
the processes $\rho \gamma^{\star} \to \pi(\rho)$ up to the two-parton
twist-3 level.

The plan of this paper is as follows. In Sec.~\ref{sec:hadron-level} we
explain the source of the factorization violation in~\cite{ChengRRA,ZhangVOR},
taking the process $\rho\gamma^{\star} \to \pi$ as an example.
The correct collinear factorization is presented in Sec.~\ref{sec:NLO}
with the help of the equation of motion for a valence quark. The above
approach applies to another considered process $\rho \gamma^{\star} \to \rho$
apparently, and can be extended to the more complicated $k_T$ factorization.
Section~\ref{sec:conclusion} contains the conclusion.

\section{Factorization violation at twist 3?}\label{sec:hadron-level}

\begin{figure}[tb]
\begin{center}
\vspace{-1.5cm}
\includegraphics[width=0.6\textwidth]{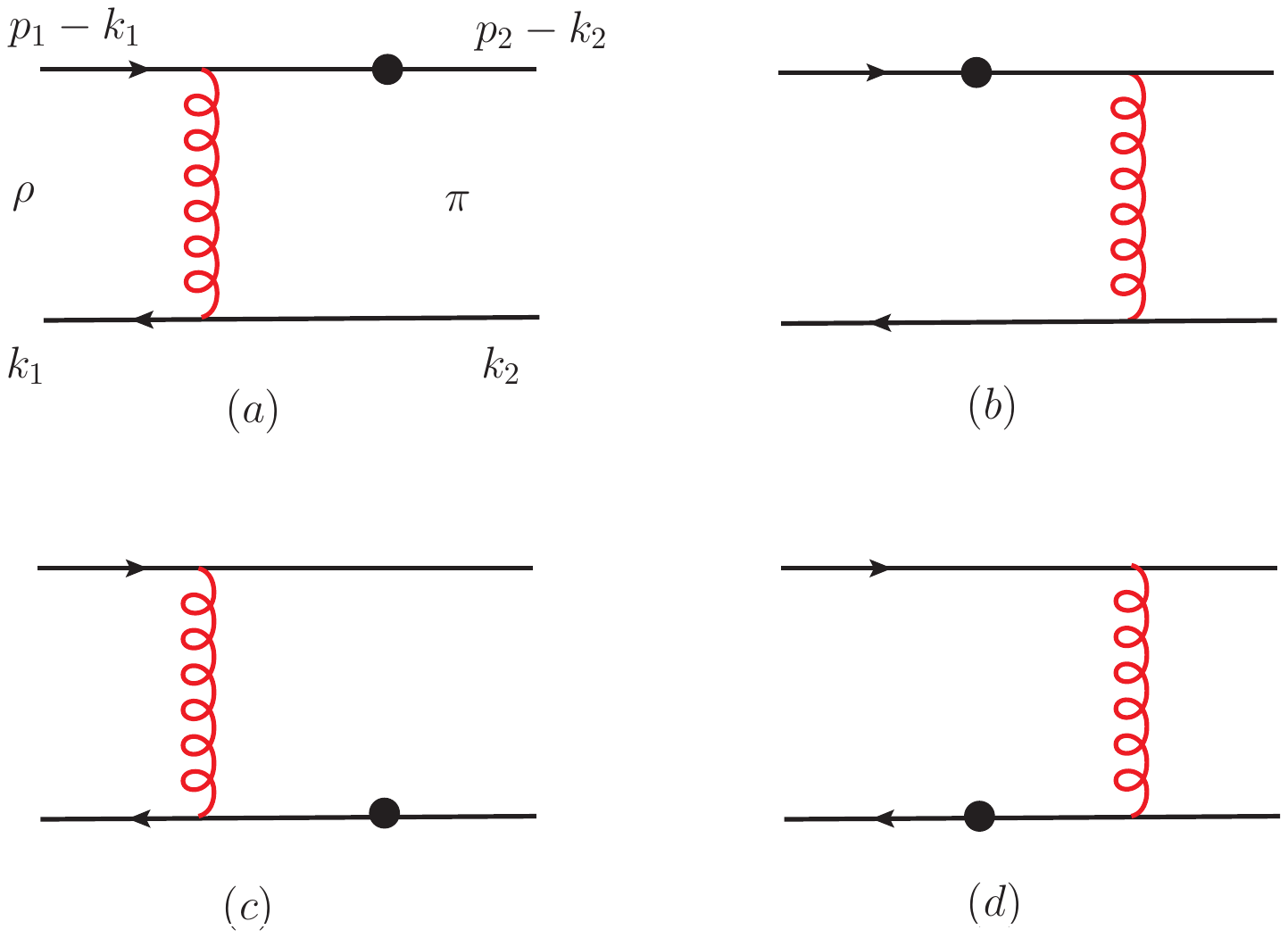}
\end{center}
\vspace{-8.5cm}
\caption{Leading-order (LO) diagrams for the scattering
$\rho \gamma^{\star} \to \pi(\rho)$.
The symbol $\bullet$ represents the virtual photon vertex.}
\label{fig:fig1}
\end{figure}

We choose the kinematic variables for the process
$\rho \gamma^{\star} \to \pi$ in the light-cone coordinates,
\beq
p_1 = \frac{Q}{\sqrt{2}}(1,0,\mathbf{0}),~~~
p_2 = \frac{Q}{\sqrt{2}}(0,1,\mathbf{0}) ,~~~k_1=x_1p_1,~~~k_2=x_2p_2\,,
\eeq
where $p_1$ ($p_2$) is the 4-momentum of the $\rho$ meson (pion),
and $k_1$ ($k_2$) is the parton momentum carried by the
antiquark in the $\rho$ meson (pion), $x_1$ and $x_2$ being the momentum fractions.
We consider the region with a large momentum transfer squared $Q^2=-(p_1-p_2)^2$,
where perturbative QCD is applicable.
To define the direction of the gauge links for the $\rho$ meson light-cone
distribution amplitudes, we introduce the dimensionless
vector $v=(0,1,\mathbf{0})$.

The LO, namely, ${\cal O}(\alpha_s)$ diagrams for the process 
$\rho \gamma^{\star} \to \pi$ are displayed in Fig.~\ref{fig:fig1}. For the purpose of
explaining the factorization violation found in~\cite{ChengRRA,ZhangVOR}, it is enough
to focus on radiative corrections to Fig.~\ref{fig:fig1}(a). Figure~\ref{fig:fig1}(a)
yields the partial $\rho \to \pi$ scattering amplitude
\beq
G^{(0)}(x_1,x_2) &=&
i e_u g^2_s C_F N_C \, \textmd{Tr} \left[
\frac{\gamma_{\nu}\,  d(k_2)\bar{u}(\bar{k}_2) \, \gamma_{\mu} \, (\psl_1-\ksl_2) \,
\gamma^{\nu} \, u(\bar{k}_1)\bar{d}(k_1)}
{(p_1-k_2)^2(k_1-k_2)^2}  \right] \,,
\label{eq:rhopi-lo-1}
\eeq
with the electric charge $e_u$ of the $u$ quark, the strong coupling $g_s$, the color
factor $C_F$, the number of colors $N_C$, the quark spinors $u$ and $d$, the momenta
$\bar{k}_1=p_1-k_1$ and $\bar{k}_2=p_2-k_2$, and the gamma matrix
$\gamma_\mu$ from the virtual photon vertex. We factorize the fermion flow by
inserting the Fierz identity
\beq
I_{ij}I_{lk} &&= \frac{1}{4}I_{ik}I_{lj}
+ \frac{1}{4}(\gamma_{5})_{ik}(\gamma_{5})_{lj}
+ \frac{1}{4}(\gamma^{\alpha})_{ik}(\gamma_{\alpha})_{lj}
+ \frac{1}{4}(\gamma_{5}\gamma^{\alpha})_{ik}(\gamma_{\alpha}\gamma_{5})_{lj}
 + \frac{1}{8}(\sigma^{\alpha\beta})_{ik}(\sigma_{\alpha\beta})_{lj} \, ,
\label{eq:fierz}
\eeq
into Eq.~(\ref{eq:rhopi-lo-1}), where
$\sigma^{\alpha \beta}=i[ \gamma^{\alpha}, \gamma^{\beta}]/2$, and different terms on
the right-hand side lead to contributions characterized by different powers in $1/Q$. The
color flow is factorized by inserting the identity
\beq
I_{ij}I_{lk} &&= \frac{1}{N_C}I_{ik}I_{lj} + 2 \sum_{c} T^c_{ik}T^c_{lj},
\label{eq:color-separate}
\eeq
with $T^c$ being a color matrix. The first term on the right-hand side is associated
with a two-parton (quark-antiquark) Fock state, which we concentrate on, and
the second term is associated with a three-parton (quark-antiquark-gluon) Fock
state.

We then arrive at the LO factorization formula in terms of the convolution in
the parton momentum fraction $\xi_{1,2}$,
\beq
G^{(0)}(x_1,x_2) &= & \int d\xi_1 d\xi_2 \,\sum_{i=1,2}
\Phi^{\pi(0)}(x_2,\xi_2) H^{(0)}_{i}(\xi_1,\xi_2)  \Phi^{\rho(0)}_{i}(x_1,\xi_1) \, ,
\label{eq:rhopi-lo-1-fact}
\eeq
between the LO meson distribution amplitudes $\Phi^{(0)}$ and the LO hard kernels $H^{(0)}$,
\beq
\Phi^{\rho(0)}_{1,2}(x_1,\xi_1)&=&\frac{1}{4}
\bar{d}(k_1)(\gamma_{\perp},\gamma_5\gamma_{\perp})u(\bar{k}_1)\delta(\xi_1-x_1)\,,
\label{rhop}\\
\Phi^{\pi(0)}(x_2,\xi_2)&=&\frac{1}{4}
\bar{u}(\bar{k}_2)\gamma^-\gamma_{5}d(k_2)\delta(\xi_2-x_2)\,,\label{pip}\\
H^{(0)}_{1,2}(\xi_1,\xi_2) &=& i e_u g^2_s C_F
\textmd{Tr} \left[\frac{\gamma_5\gamma^+\gamma_{\mu}(\psl_1-\xi_2\psl_2)\gamma^{\nu}
(\gamma_{\perp},\gamma_{\perp}\gamma_5)
\gamma_{\nu}}{(p_1-\xi_2 p_2)^2(\xi_1 p_1-\xi_2p_2)^2}\right]\,.
\label{hard}
\eeq
The spin projectors $\gamma_{\perp}$, $\gamma_{\perp}\gamma_5$, and
$\gamma_5\gamma^+$, which give the nonvanishing hard kernels
$H^{(0)}_{1,2}(\xi_1,\xi_2)$, come from Eq.~(\ref{eq:fierz}). Since only a
transversely polarized $\rho$ meson can transit into a pseudoscalar pion,
the spin projectors involving $\gamma_\perp$ are relevant. Both $\Phi^{\rho}_{1}$
and $\Phi^{\rho}_{2}$, defined by $\gamma_{\perp}$ and $\gamma_5\gamma_{\perp}$,
respectively, represent the two-parton twist-3 $\rho$ meson distribution
amplitudes. The pion distribution amplitude $\Phi^{\pi}$ defined by
$\gamma^-\gamma_{5}$ is of twist 2. Without collinear gluon exchanges at LO,
the modified momentum fractions are equal to the initial fractions in the
external mesons, as indicated by the delta functions $\delta(\xi-x)$
in Eqs.~(\ref{rhop}) and (\ref{pip}).

A crucial step to prove the factorization theorem is to explore
infrared divergences in radiative corrections, and to examine whether
soft divergences cancel and collinear divergences are absorbed completely
into hadronic matrix elements. The former are generated when the
momentum of a radiative gluon, exchanged between two on-shell particles,
vanishes like $l \sim (\Lambda,\Lambda,\Lambda)$, with $\Lambda$ denoting
a small scale. The latter appear, when a radiative gluon is collimated to
on-shell massless particles. As a gluon is aligned with the initial $\rho$
meson, its momentum scales like $l \sim (Q,\Lambda^2/Q,\Lambda)$. The
factorization of collinear divergences is first verified at NLO, and then
generalized to all orders by the induction~\cite{LiHH,NagashimaIW}.
As stated in the introduction, the NLO analysis of the infrared divergences
in the scattering $\rho\gamma^{\star} \to \pi$ was performed by adding radiative
gluons to the hard kernels in Eq.~(\ref{hard})~\cite{ChengRRA,ZhangVOR}, instead
of to the scattering amplitude in Eq.~(\ref{eq:rhopi-lo-1}). Since a hard kernel
is not a physical quantity, the matching between the full QCD and the effective
theory for infrared physics may not be implemented correctly due to wrong power
counting for collinear divergences.

The NLO corrections to Fig.~\ref{fig:fig1}(a) with the additional gluon
radiated from the initial $\rho$ meson are displayed in Fig.~\ref{fig:fig2}.
Take Fig.~\ref{fig:fig2}(d) with a triple-gluon vertex,
sandwiched by $\gamma_{\perp}$ or $\gamma_{\perp}\gamma_5$ from the
$\rho$ meson side and $\gamma_5\gamma^+$ from the pion side, as an example.
It contains the Feynman rule
\beq
F_{\alpha\beta\gamma}=g_{\alpha\beta}(2k_2-2k_1+l)_{\gamma}
+g_{\gamma\alpha}(k_1-k_2 +l)_{\beta}+g_{\beta\gamma}(k_1-k_2-2l)_{\alpha}\, ,
\label{eq:triple-gluon-vertex}
\eeq
where the indices $\alpha$, $\beta$ and $\gamma$ are labeled in the diagram.
In the collinear region with the loop momentum $l$ being almost parallel
to $p_1$, the leading contribution arises from the gamma matrix
$\gamma^\gamma=\gamma^+$, because the adjacent quark propagator
is proportional to $\psl_1-\ksl_1+\lsl \propto \gamma^-$. The first
term in Eq.~(\ref{eq:triple-gluon-vertex}) yields the expected
factorization of a collinear gluon: the metric tensor $g_{\alpha\beta}$
is identified as the one in the LO hard kernel; the term
$2k_{2\gamma}=2k_2^-$, picked up by the vertex
$\gamma^\gamma=\gamma^+$, facilitates the splitting of the two
gluon propagators,
\beq
\frac{2k_{2\gamma}}{(k_1-k_2-l)^2(k_1-k_2)^2}
\approx \left[ \frac{1}{(k_1-k_2)^2} - \frac{1}{(k_1-k_2-l)^2} \right]
\frac{v_\gamma}{v \cdot l}.
\label{eikonal}
\eeq
The first (second) term in the above square brackets corresponds to the
LO hard kernel without (with) the loop momentum flowing through it.
The eikonal vertex $v_\gamma$ and the eikonal propagator $1/v \cdot l$
are the Feynman rules associated with the gauge links along the direction
$v$, which are necessary for defining gauge-invariant nonlocal
hadronic matrix elements. Hence, the collinear divergence from
the first term in Eq.~(\ref{eq:triple-gluon-vertex}) contributes to the
NLO two-parton twist-3 $\rho$ meson distribution amplitudes
$\Phi^{\rho(1)}_{1,2}$.

Nevertheless, the second term in Eq.~(\ref{eq:triple-gluon-vertex}) also
gives rise to a collinear divergence as $l$ is parallel to $p_1$, which does
not respect the factorization. The gamma matrix $\gamma^\alpha$ can be set
to $\gamma^-$ owing to the spin projectors $\gamma_{\perp}$ or
$\gamma_{\perp}\gamma_5$ from the $\rho$ meson side, and $\gamma_5\gamma^+$ from
the pion side. Then $\gamma^\alpha=\gamma^-$ and $\gamma^\gamma=\gamma^+$ for
the collinear gluon contract to the tensor
$g_{\gamma\alpha}$. The gamma matrix $\gamma^\beta$ is chosen as $\gamma^+$,
also because of the adjacent quark propagator proportional to
$\psl_1-\ksl_1+\lsl \propto \gamma^-$,
which picks up the nonvanishing component $-k_{2\beta}=-k_2^-$. The above
configuration produces the residual collinear divergence, which cannot be
absorbed into $\Phi^{\rho(1)}_{1,2}$ having been defined by the contribution from
the first term in Eq.~(\ref{eq:triple-gluon-vertex}).
The third term in Eq.~(\ref{eq:triple-gluon-vertex}) does not produce
a collinear divergence, since $\gamma^\gamma=\gamma^+$ requires
$\gamma^\beta=\gamma^-$ through $g_{\beta\gamma}$, which then suppresses
the adjacent quark propagator proportional to $\psl_1-\ksl_1+\lsl \propto \gamma^-$.
As elaborated comprehensively in Ref.~\cite{ZhangVOR}, the other
triple-gluon diagrams, sandwiched by the various twist-3 spin projectors,
generate the similar residual collinear divergences. Note that this
source of factorization violation does not exist, as the NLO diagrams are sandwiched
only by the twist-2 spin projectors: the twist-2 spin projector $\gamma^-$
from the initial $\rho$ meson would suppress the violation source from
$\gamma^\alpha=\gamma^-$.

\section{Proof of the collinear factorization}\label{sec:NLO}

As postulated in the introduction, the correct procedure to examine
the factorization starts with analyzing infrared structures of
higher-order scattering amplitudes, for which the equations of motion
obeyed by valence quarks can be applied~\cite{LiHH,NagashimaIW}.
In this section we demonstrate, with
the help of the equations of motion, the factorization of the collinear
divergences in the scattering $\rho \gamma^{\star} \to \pi$ at NLO,
focusing on those associated with the initial $\rho$ meson. The
approach in~\cite{ChengRRA,ZhangVOR} does not cause a problem at the
leading-twist accuracy, because the effect of the twist-2 spin projectors,
as mentioned at the end of the previous section, is equivalent to the
equations of motion for valence quarks.

\begin{figure}
\vspace{0.5cm}
\begin{center}
\includegraphics[width=0.8\textwidth]{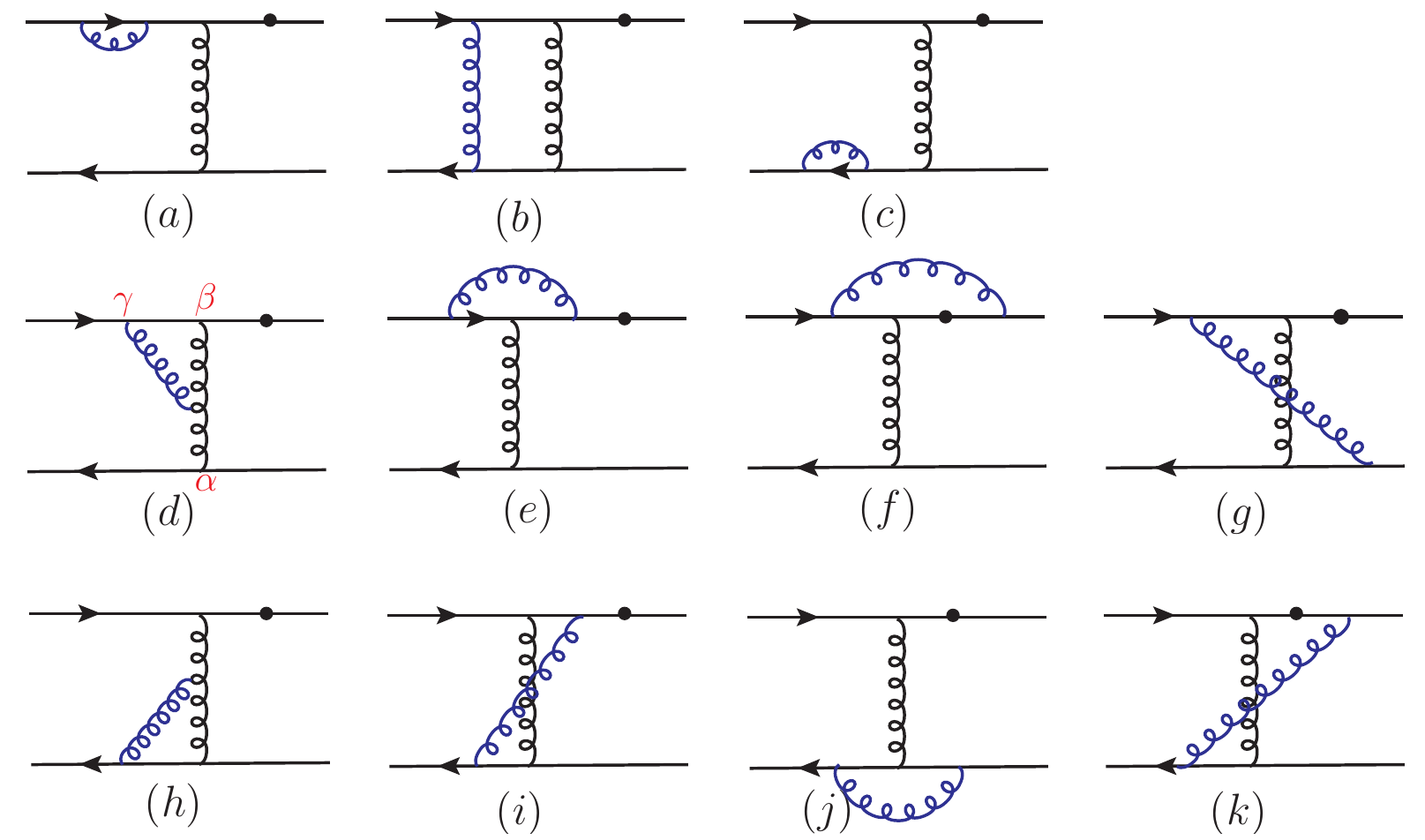}
\end{center}
\vspace{-0.5cm}
\caption{NLO corrections to Fig.~\ref{fig:fig1}(a) with the
gluon radiated from the initial $\rho$ meson.}
\label{fig:fig2}
\end{figure}

An energetic valence quark of momentum $k=(k^+,k^-,\mathbf{k_T})$,
as an on-shell parton with $k^-=k_T^2/(2k^+)$, satisfies the equation of motion
\beq
\ksl u(k) = \left(k^{+} \gamma^{-} + k^- \gamma^{+}
-\mathbf{k_T}\cdot \mathbf{\gamma_\perp} \right)\, u(k)=0\,,
\label{quark-field-decomposition}
\eeq
when the three-parton terms are neglected. The above decomposition
leads to the power counting
\beq
\gamma^- u(k) \sim
\calo\left(\frac{\Lambda}{Q}\right)\gamma_\perp u(k)\,,
\label{power}
\eeq
for $k^+\sim \calo(Q)$ and $k_T\sim \calo(\Lambda)$. That is, the product
$\gamma^- u(k)$ is suppressed by a power of $1/Q$ relative to $\gamma_\perp u(k)$.

For the first term in Eq.~(\ref{eq:triple-gluon-vertex}), $\gamma^\alpha$ can
be set to $\gamma_\perp$ ($\gamma^\beta$ also takes $\gamma_\perp$ due to
the tensor $g_{\alpha\beta}$). As explained in the previous section, the source
of the factorization violations comes from $\gamma^\alpha=\gamma^-$. By considering
the infrared structure of the scattering amplitude, instead of the hard kernels,
the above  $\gamma_\perp$ and $\gamma^-$ attach to the spinor of the valence
antiquark. The power counting in Eq.~(\ref{power}) then implies that the contribution
from the latter is down by a power of $1/Q$ compared to the contribution from the former,
which respects the factorization at twist 3.
Therefore, the violation source should be dropped
at the twist-3 accuracy, and the proof of the factorization theorem at
twist 3 based on the equations of motion in~\cite{LiHH,NagashimaIW}
is verified. After absorbing the collinear divergences
into the two-parton twist-3 $\rho$ meson distribution amplitudes,
we insert the Fierz identity to factorize the fermion flow between the
NLO distribution amplitudes and the remaining LO hard kernels.
The NLO factorization of the collinear gluons emitted by the outgoing
valence quark and antiquark of the pion at the twist-2 level is the same as
in~\cite{LiHH}, so we do not touch it here.

As to the soft divergences, it is straightforward to show that they
cancel each other among the reducible diagrams
Figs.~\ref{fig:fig2}(a)$-$\ref{fig:fig2}(c), between the irreducible diagrams
Figs.~\ref{fig:fig2}(f) and \ref{fig:fig2}(g), as well as between
Figs.~\ref{fig:fig2}(j) and \ref{fig:fig2}(k)~\cite{LiHH,NagashimaIW}.
This soft cancellation can be understood via the color-transparency argument
for an energetic $\rho$ meson. In other words, a soft gluon extends over a huge
space-time, so it cannot resolve the color structure of the $\rho$ meson.
Figures~\ref{fig:fig2}(d), \ref{fig:fig2}(e), \ref{fig:fig2}(h), and \ref{fig:fig2}(i)
do not contain soft divergences, since the radiative gluons attach to
the LO hard particles.

The proof of the collinear factorization to all orders in the strong
coupling follows the induction method based on the Ward
identity~\cite{LiHH,NagashimaIW}: the eikonal approximation
holds for every internal particle line which a collinear gluon attaches
to (in the absence of the Glauber divergences); the summation
over all possible attachments of a collinear gluon
to internal lines leads to its factorization in color space;
at last, the induction extends the factorization from
the assumed order to the next higher order. The above steps
complete the all-order proof of the collinear factorization for
$\rho\gamma^{\star} \to \pi$ up to the two-parton twist-3 accuracy, and
we arrive at the definition of the two-parton twist-3 $\rho$ meson
distribution amplitudes
\beq
\Phi^{\rho}_{1,2}(\xi_1,x_1)
= \int \frac{dy^-}{2\pi} e^{i\xi_1p_1^+y^-}
\langle 0 \vert \bar{d}(y^-) \, (\gamma_{\perp},\gamma_5\gamma_{\perp}) W_{v}(y^-) \, u(0)
\vert u(\bar{k}_1)\bar{d}(k_1)\rangle \, ,
\label{eq:Wilson-line-formular-rho}
\eeq
where $W_{v}$ is a path-ordered exponential function
\beq
W_{v}(y^-) = \calp \,\, \textmd{Exp} \left[
-i g_s \int_{0}^{y^-} dz v \cdot A(zv) \right] \, .
\label{eq:Wilson-line-rho}
\eeq
It is easy to see that the above gauge links produce the Feynman rules
described by the eikonal vertex and propagator in Eq.~(\ref{eikonal}).
The discussion of the scattering  $\rho \gamma^{\star} \to \rho$, i.e., the $\rho$
meson electromagnetic form factor, is basically the same as of
$\rho \gamma^{\star} \to \pi$ before inserting the appropriate spin projectors
associated with the final-state meson, so we do not repeat it in this work.
To derive the physical two-parton twist-3 $\rho$ meson distribution amplitudes
$\Phi^{\rho}_{1,2}(\xi_1)$, we simply replace the Fock state
$\vert u(\bar{k}_1)\bar{d}(k_1)\rangle$ in Eq.~(\ref{eq:Wilson-line-formular-rho})
by the $\rho$ meson state $\vert\rho(p_1,\epsilon_1) \rangle$, with $\epsilon_1$
being the polarization vector.

The $k_T$ factorization is more apposite to QCD processes dominated by
contributions from small parton momenta. For its proof
as an extension of the collinear factorization, we retain the dependence
of parton transverse momenta $k_T$ in hard kernels. The factorization
of collinear divergences in radiative corrections,
which are then regularized by the small parton off-shellness into
$\ln k_T^2$, gives TMD hadron wave functions.
A TMD wave function, collecting the collinear logarithm $\ln k_T^2$ to all orders,
describes parton distributions in both longitudinal and transverse momenta
inside a hadron. Note that the $k_T$ terms appearing in numerators of internal
particle propagators are still dropped~\cite{NagashimaIA}, whose contribution is
supposed to be combined with that from the three-parton Fock state \cite{ChenPN}
to form a gauge-invariant three-parton wave function. Moreover, neglecting
$k_T$ in numerators allows the eikonal approximation to hold for collinear
gluons, so that the gauge links, which guarantee the gauge invariance of a TMD
wave function, can be constructed. Following the similar procedure
in~\cite{NagashimaIA}, the $k_T$ factorization for a high-energy QCD process
can be proved, despite of the above distinct features, to the two-parton
twist-3 accuracy.

It should be stressed that we can no longer drop the product $\gamma^- u(k)$, i.e.,
the residual collinear divergences at the twist-4 accuracy. Hence, we speculate
that the factorization of the $\rho\gamma^{\star} \to \pi$ scattering amplitude
into a convolution involving the twist-2 and twist-4 meson distribution amplitudes
for the initial and final states separately may break down. We do not pursue
either the factorization for the cases with both the initial and final states
taking the twist-3 meson distribution amplitudes, which exhibit a power-law
behavior the same as in the combination of the twist-2 and twist-4 meson
distribution amplitudes.

\section{Conclusion}\label{sec:conclusion}

In this paper we have elaborated how the residual collinear divergences,
which violate the factorization of the exclusive processes
$\rho\gamma^{\star} \to \pi(\rho)$ at the two-parton twist-3 level,
appear in the proof of~\cite{ChengRRA,ZhangVOR}. They are attributed to the
improper factorization of the fermion flow in the scattering amplitudes
before absorbing the collinear divergences into the hadronic matrix elements.
Namely, the NLO analysis of the infrared divergences was
performed by adding radiative gluons to the unphysical hard kernels,
instead of to the scattering amplitudes. This is the reason why the
residual collinear divergences survive under the wrong power counting,
and break the factorization theorem.
We have proposed to study the infrared structure of the higher-order scattering
amplitudes first, such that the equations of motion for valence quarks can
be applied. With the correct power counting based on the equations of motion, it
has been shown that the residual collinear divergences are in fact
power suppressed, and negligible at the twist-3 accuracy.
The Fierz identity is inserted into the scattering amplitudes to factorize
the fermion flow, after the collinear divergence has been absorbed into the
meson distribution amplitudes completely. The rest of the proof follows the
steps outlined in~\cite{LiHH,NagashimaIW}, which is then extended to all
orders by the induction. The above procedure works for the more complicated $k_T$
factorization, and for the factorization of other high-energy exclusive
processes, including $B$ meson transition form factors.

\section*{Acknowledgments}
We thank Yue-Long Shen and Wei Wang for useful discussions.
This work was supported in part by the DFG Research Unit FOR 1873 Quark
Flavor Physics and Effective Theories under Contract No. KH 205/2-2, by
the Ministry of Science and Technology of R.O.C. under
Grant No. MOST-104-2112-M-001-037-MY3, and by the National Natural Science Foundation
of China under Grant No. 11235005.

\end{document}